\documentclass[conference]{IEEEtran}
\IEEEoverridecommandlockouts
\usepackage{amsmath,amssymb,amsfonts}
\usepackage{algorithmic}
\usepackage{graphicx}
\usepackage{textcomp}
\usepackage{relsize}
\usepackage{svg}
\usepackage{tabularx}
\usepackage{array}
\usepackage{subfig}

\usepackage{xcolor}
\usepackage[hyphens]{url}
\def\BibTeX{{\rm B\kern-.05em{\sc i\kern-.025em b}\kern-.08em
    T\kern-.1667em\lower.7ex\hbox{E}\kern-.125emX}}
    
\usepackage{biblatex}
\begin{document}

\title{Click-Through Rate Prediction Using Graph Neural Networks and Online Learning}

\author{\IEEEauthorblockN{Farzaneh Rajabi}
\IEEEauthorblockA{\textit{Stanford University} \\
frajabi@stanford.edu}
\and
\IEEEauthorblockN{Jack Siyuan He}
\IEEEauthorblockA{\textit{Google LLC} \\
siyuanhe@stanford.edu}
}

\maketitle
\thispagestyle{plain}
\pagestyle{plain}

\section{Introduction}

Recommendation systems have been extensively studied by many literature in the past and are ubiquitous in online advertisement, shopping industry/e-commerce, query suggestions in search engines, and friend recommendation in social networks. Moreover, restaurant/music/product/movie/news/app recommendations are only a few of the applications of a recommender system. A small percent improvement on the CTR prediction accuracy has been mentioned to add millions of dollars of revenue to the advertisement industry.
Click-Through-Rate (CTR) prediction is a special version of recommender system in which the goal is predicting whether or not a user is going to click on a recommended item. 

A content-based recommendation approach takes into account the past history of the user's behavior, i.e. the recommended products and the users reaction to them. So, a personalized model that recommends the right item to the right user at the right time is the key to building such a model. On the other hand, the so-called collaborative filtering approach incorporates the click history of the users who are very similar to a particular user, thereby helping the recommender to come up with a more confident prediction for that particular user by leveraging the wider knowledge of users who share their taste in a connected network of users. 

Using any of these two approaches, an accurate prediction demands framing the problem in such a way that the nonlinear high-order interaction between features of the dataset are taken into account. Graph neural Networks are proposed to assist with this purpose. However, even by considering the high-order feature interactions, framing the recommendation as a static procedure and ignoring the dynamic interactive nature between users and the recommender systems is yet a limiting factor, resulting in a discrepancy between offline metrics for supervised learning and the online performance of the newly proposed models. As a solution, reinforcement learning treats the recommendation systems as a sequential decision making process that models the interactions between the user and the recommender engine as a dynamic adaptation scheme. That is, the agent observes the environment and acts upon that in order to change its state towards better states (states with higher rewards).

In this project, we are interested in building a CTR predictor using Graph Neural Networks complemented by an online learning algorithm that models such dynamic interactions. By framing the problem as a binary classification task, we have evaluated this system both on the offline models (GNN, Deep Factorization Machines) with test-AUC of 0.7417 and on the online learning model with test-AUC of 0.7585 using a sub-sampled version of Criteo public dataset consisting of 10,000 data points.

\section{Literature Review}

Machine learning based CTR predictor has been thoroughly discussed in the past literature. These models started with content-based to recommend items by considering the content
similarity between items.  Later, collaborative filtering was introduced with the rationale behind that the users with similar behaviors tend to prefer the same items, and the items consumed by similar users tend to have the same rating. However, conventional CF-based methods tend to suffer from the data scarcity, and the similarity calculated from sparse data can be very unreliable. The simplest and somewhat trivial model involves Logistic Regression (LR) \cite{Chapelle_lr} that can only model first-order interactions by linearly combining extracted features. Such a model works if features are linearly dependent. However, when features have second-order interactions with each other, this model may fail to provide accurate results. Naturally, we would like to create joint features using two or more raw features, thereby taking the correlation between the features into account. However, this leads to a quadratic explosion of feature space, rendering learning task infeasible. 

Then we found that Factorization Machines (FM) \cite{rendle2010factorization} came to rescue. FM is able to represent second-order features as factorized parameters whose space grows linearly with respect to the total number of features. Using FM, researchers were able to improve the CTR prediction where second-order feature interactions were present \cite{juan_fm_ctr}. 

However, natural features may interact beyond second-order, making FM not sufficient for certain use cases. The initial approach taking by the researchers was to put those features into a deep neural network (DNN). Past works like DeepFM \cite{guo2017deepfm} were able to successfully demonstrate CTR prediction improvement over FM. In order to de-bias the system towards the low- or high- order feature interactions, DeepFM proposes an end-to-end learning model that emphasizes on both low- and high order feature interactions. They combine the power of factorization machines for recommendation and deep learning for feature learning in a new neural network architecture. The FM part models the linear (1st order) and pairwise (2nd order) feature interactions as inner product of respective feature latent vectors. Whereas, the deep component is a feed-forward neural network,
which is used to learn high-order feature interactions. Compared to the Wide \& Deep model from Google \cite{cheng2016wide}, DeepFM has a shared input to its “wide” and “deep” parts, with no need of feature engineering besides raw features, making them a powerful practical model.

While being better than previous models, DNN-based models use bitwise factors and the learning results are internal to the models themselves and therefore they lack explainability. This can be bad for business critical models that require logical explanation from their users. Moreover, these implicit models do not allow manual input of existing models that could speed-up the training process or improve the metrics. As the result, researchers turned to more explicit models such as Deep\&Cross \cite{deepAndCross2017} that takes outer product of features at bit level. 

Very recently, a graph-based neural network model, Fi-GNN \cite{Li_2019}, was developed to explicitly model relations among features in a graph. In this work, the multi-field features are represented in a graph structure, where each node corresponds to a feature filed, and different fields interact with each other through edges. Therefore, the task of modeling feature interactions is converted to modeling node interactions on the corresponding graph.
Generally, nodes in GNNs interact with neighbors by aggregating information from neighborhoods and updating their hidden states. There have been many variants of GNN with various kinds of aggregators and updaters. For instance, Gated Graph Neural Networks use GRUs as updater.  Graph Convolutional Networks utilize the convolutional aggregators. Graph Attention Networks incorporate the attention mechanism into the propagation step. In \cite{Li_2019}, the input sparse m-field feature vector is first mapped into sparse one-hot
embedding vectors and then embedded to dense field embedding vectors via the embedding layer and the multi-head self-attention layer. The field embedding vectors are then represented as a feature graph, where each node corresponds to a feature field and different feature fields can interact through edges. An attention scoring layer is applied on the output of Fi-GNN to estimate the CTR. A log-loss function as below is used as the loss function,
\begin{align}
  \mathcal{L} = -\dfrac{1}{N}\mathlarger{\sum}_{i=1}^N \bigg(y_i log(\hat{y}_i) + (1-y_i) log(1-\hat{y}_i)\bigg)  
\end{align}
Also, the edge-wise interactions are achieved via attentional edge weights and edge-wise transformations, and extra residual connection along with GRU is introduced to update states, which can help regain the low-order information.
They have reported superior accuracy boost over the sate-of-the-art models for CTR predictions evaluated on the public benchmark dataset (in terms of their evaluation metrics: AUC, Logloss(cross entropy) and Relative Improvement (improvement with respect to AUC, Logloss)).

That being said, yet most CTR predictors we found so far are offline models that require re-training when new data is available and cannot improvise to explore different recommendation strategies. There is a big issue regarding offline models like such. Since data is created at enormous rate in today's world, we may not have enough data or time to re-train our offline model. For example, a natural disaster like COVID-19 may suddenly change users' preference without giving enough data and time to retrain the models. Moreover, before retraining the model, offline models will predict CTR using previously trained parameters, resulting in an unsatisfactory period between the change and new model update. Moreover, offline models often fail to address the internal state of users. For example, a user who has already bought a washing machine would most likely not buy another one. In this case, stateful models came to rescue. We read about works that use adversarial sampling \cite{adversalNetworkRL2019} for CTR prediction. Although such works exploited real-time feedback into their pipeline, it requires prior knowledge of the model itself and this knowledge about the model cannot be updated in real-time. 

Subsequently, Reinforcement Learning (RL) models were introduced. Past work \cite{deepRLNewRecommendation2018} and \cite{liu2018deep} demonstrated the ability to use RL for news article recommendation and movie recommendation, respectively. In the former literature, they use a combination of online-offline models to predict users' likelihood of clicking on a news article. Updates generated by RL is incorporated into the model through a combination of minor and major updates to avoid frequent disturbances to the model. A similar approach can be applied in an advertisement CTR prediction task.

Model-based RL techniques are proposed to model recommendation procedure, such as POMDP \cite{shani2005mdp} and Q-learning \cite{taghipour2008hybrid}. However, these methods are inapplicable to complicated recommendation scenarios when the number of candidate items is large, because a time-consuming dynamic programming step is required to update the model. Later, model-free RL techniques were introduced, divided into two categories: value-based and policy-based. Value-based approaches compute Q-values of all available actions for a given state and the one with the maximum Q-value is selected as the best action.
However, due to the evaluation on overall actions, the approaches may become very inefficient if the action space is too large. Policy-based approaches generate a continuous parameter vector as the representation of an action, which can be utilized in generating the
recommendation and updating the Q-value evaluator. The inefficiency drawbacks
can be overcome due to the continuous representations. Yet, these models have one common limitation: the user state is learnt via a conventional fully connected neural network, which does not explicitly and carefully model the interactions between users
and items.

The approach presented in \cite{liu2018deep} overcomes this constraint using an “Actor-Critic” type framework incorporated with a state representation module, which explicitly models the complex dynamic user-item interactions, thereby treating recommendation as a sequential decision making problem, in which the recommender (i.e., agent) interacts with users (i.e., environment) to suggest a list of items sequentially over the timesteps, by maximizing the cumulative rewards of the whole recommendation procedure.
The former literature is an interesting model that could be applied on the Criteo dataset. However, given our timeline for this class, we are inspired by the approach in \cite{deepRLNewRecommendation2018} and will follow a similar approach.

\section{Dataset}

We used the Criteo dataset\footnote{\url{https://www.kaggle.com/c/criteo-display-ad-challenge/data}} originally released by Criteo Labs as part of the Criteo Display Ads Challenge. The benefits of using such a dataset are:
\begin{itemize}
    \item It is freely available online for immediate download
    \item The data format is very standardized 
    \item It's one of the few publicly available benchmark datasets, used by many researchers including \cite{Li_2019}, meaning that we can compare our results with theirs
\end{itemize}
The dataset consists of a portion of Criteo's traffic over a 7 day period, which is split into training and test dataset. Each row represents an AD shown through Criteo's display ADs network where the first column represents whether it was clicked (label 1) or not (label 0) by that specific user. There are 39 anonymized feature fields; the first 13 columns are integer features, followed by 26 categorical feature columns. The training dataset contains about 45 million rows and the test dataset contains about 6 million rows. Due to the limiting size of the entire dataset, we used a small portion of it consisting of 10000 user's click records, and we divided it into 80\% train, 10\% validation and 10\% test portions.

The data comes in CSV format, which needs to be converted into NumPy format first. Then the data needs to be properly scaled if log or quadratic feature values are needed. AutoInt has provided an open source implementation of the above prepossessing steps\footnote{\url{https://github.com/DeepGraphLearning/RecommenderSystems/tree/master/featureRec}} \cite{song2019autoint}. In this process, the data is also segmented into 10 smaller parts so that it fits into a normal computer's memory. 

Unfortunately, due to the time constraints, we could not investigate the data-imbalance for the data portion that we used. However, we are aware that investigating the data in terms of imbalance is the key to a reliable accuracy. That is, if the number of examples with positive and negative labels are not almost equally distributed, we have to do either majority-class down-sampling or minority-class up-sampling as a potential solution to resolve this issue. SMOTEBoost and RUSBoost have been proposed to alleviate the class imbalance problem \cite{chawla2003smoteboost}, \cite{seiffert2009rusboost}.

That being said, we are suspect that such an issue could be the underlying reason for not getting a higher AUC from our implemented model (and also, the reason behind getting a lower accuracy than what \cite{Li_2019} has reported). This came into our mind after running the experiments and looking at the evaluation metrics, but is definitely something of interest to inspect, and see whether the performance improves or not.

\section{Baseline}

As for our baseline model, we have used the Fi-GNN's open source implementation. The Feature-interaction Graph Neural Network implementation has modeled the low- and high- order feature interactions by modeling the features as nodes and interactions as edges. Therefore, we used it as a pre-training step for our online learning algorithm, and ran it with our parameters. This GNN model has been described thoroughly in the literature review section. 

However, initially, after getting Fi-GNN's code from GitHub, we obtained a local test AUC of 0.5433 (meaning that model has almost no class separation capacity) and a test LogLoss of 0.6999 as mentioned in our progress report. This is very far away from their claimed AUC of around 0.8061 and LogLoss of around 0.4455 in their paper \cite{Li_2019}. We initially suspected that this discrepancy was due to our usage of a sub-sampled smaller version of the Criteo dataset. This is partially true. In their GitHub repository, they included a sub-sampled version of the Criteo dataset that includes 10,000 data points, whereas the full dataset has 45 million data points. However, this was not the main reason for the difference. 

Through our exploration, we found that Fi-GNN first splits the dataset into 10 equal partitions. For the 10,000 data point sample, each partition contains 1,000 data points. During training phase, their model loads each partition into memory and train on its data in batches. There is one problem here: their default batch size was set to 1024. This batch size is OK if the number of data points in each data partition is large. For example, for the 45 million data points full dataset, each partition would have 4.5 million data points and training in 1024 data points per batch would still result in around 4,500 training iterations per data partition. However, for the sub-sampled dataset, there is only 1000 data points per data partition. As the result, a batch size of 1024 would result in only one training iteration per data partition. Since Fi-GNN train its model on only eight data partitions, the total number of training iteration when batch size is 1024 is only 80. The lack of sufficient training iterations make the model not fully updated and therefore performs very poorly. 

After realizing that, we changed the batch size to 100 and the result improved significantly. With 100 data points per batch, each 1000 data partition would trigger 10 training iterations, summing up to 800 training iterations for all 8 training data partitions. With this change, we were able to get a test AUC of 0.7417 and test log loss of 0.4687 using only the sub-sampled dataset. We took this as our new baseline when evaluating our online learning models. 

On a side note, we have also tried to use Deep-Factorization Machine implementation as our second baseline. The main repository contains a lot of models including Deep and Cross Neural Networks, however, we have only employed whatever is relevant to the DeepFM model. This model is composed of a deep component and a FM component that share the same input. The FM component is a factorization machine, which is proposed in \cite{rendle2010factorization}. The deep component is a feed-forward neural network that models the higher-order (higher than second-order) feature interactions, resulting in a higher accuracy. The FM component and deep component share the same feature embedding, which brings two important benefits: It learns both low- and high-order feature interactions from raw features. Also, there is no need for expertise feature engineering of the input, as required in \cite{cheng2016wide}.
We got test-AUC of 0.6428 and test-Logloss of 1.5371 from this model without any hyper-parameter tuning. 

\section{Main approach}

Our goal in this project is not to implement a new CTR predictor, but to seek improvements on an existing CTR predictor using some online learning algorithm. Therefore, we took Fi-GNN as our base implementation. By default, this model is only trained offline and cannot be updated during online predictions. In our experiment configuration, we made several improvements to this model to make it an online model that can be updated with more inputs. 

Ideally, we want to be able to update the weight vector in the model's neural network using Reinforcement Learning algorithms like Q-Learning. However, this model has a complicated seven layer graphical neural network that is buried in a TensorFlow session. Getting the weights out and updating them via Q-Learning will be extremely challenging and hard to achieve in the span of this quarter-long class project. As an alternative, we opt for online training, in which we feed test input and ground truth back to the Fi-GNN model and train it on the fly. 

\begin{figure}[htbp]
\centering 
\includegraphics[width=\columnwidth]{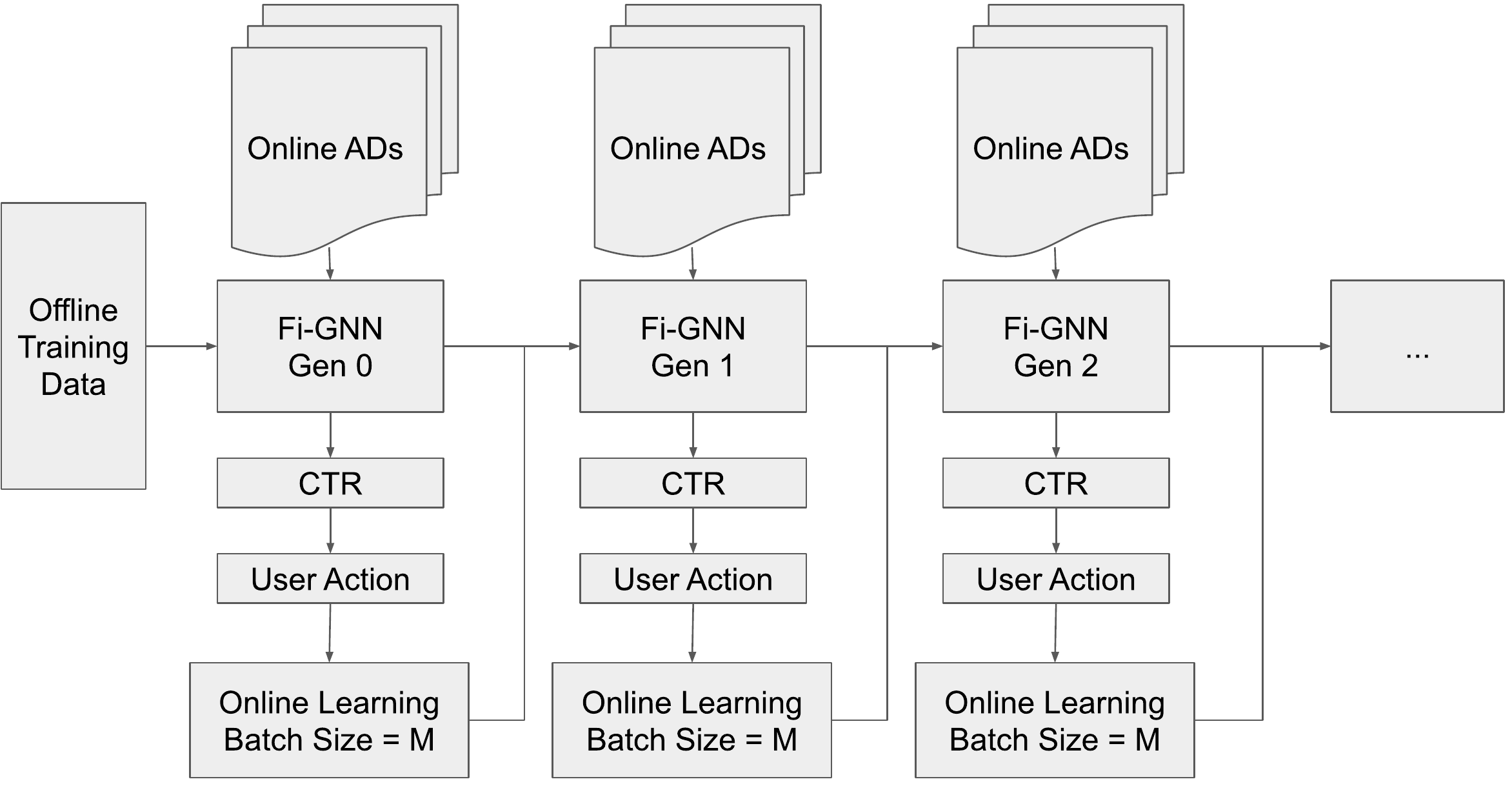} 
\caption{Pipeline Design}
\label{fig:pipeline}
\end{figure}

In particular, as shown in our pipeline design in Figure \ref{fig:pipeline}, we firstly train Fi-GNN using eight partition offline dataset of 8,000 data points. The Fi-GNN model is then validated against another 1,000 data points. Then we throw in the test dataset to simulate a real world ADs serving environment. In particular we let the current Fi-GNN model predict the CTR given a test AD input. Then after M test AD inputs, we accumulate all test ADs input in a batch and use them to re-train the Fi-GNN model so that the model evolve into a new generation. We then use the model to predict CTR for future ADs input. This cycle continues until we exhaust all online ADs input. In our test environment, we only have 1,000 online ADs input. In reality, there should be infinite test ADs input when we try to serve user with some ADs from our database. 

We opt for this batched online training approach as we were inspired by a similar work using Reinforcement Learning instead \cite{deepRLNewRecommendation2018}. After getting the pipeline running, we started tweaking the online learning batch size, M, and observing how our evaluation metrics change with respect to changes in batch size, M, especially the Area Under Curve (AUC) metric that would be explained in the next section.  

\section{Evaluation Metric}

The main evaluation metrics we used were similar to the ones in relevant literature; AUC (Area Under the ROC curve) \cite{FAWCETT2006861} and LogLoss under the same definition as Fi-GNN \cite{Li_2019}. In particular:
\begin{itemize}
    \item \textbf{AUC} measures the probability that a clicked item will be chosen over a randomly selected unclicked one. Therefore, the higher the AUC, the better.  It tells how much model is capable of distinguishing between classes.
    \item \textbf{LogLoss} measures the distance between the predicted CTR and the true label for each AD, the lower the better.
    \item \textbf{Precision} is the ratio (true positive)/(true positive + false positive). It's the ability of the classifier not to label as positive a sample that is negative. It ranges from 0 (worst) to 1 (best), and is a measure of the models' relevancy.
    \item \textbf{Recall} is the ratio (true positive)/(true positive + false negative). It's the ability of the classifier to find all the positive samples. It ranges from 0 (worst) to 1 (best), is a measure of the model's completeness.
    \item \textbf{F1-score} is the weighted average of precision and recall, where it has its best value at 1 and worst value at 0. So, F1= 2*(precision*recall)/(precision + recall). As the output from our model ($\hat{y}$) is in fact a probability, while the label vector $y$ has only 0 and 1s, we have plotted this measure for different thresholds above which we take $\hat{y}$ to be 1, and otherwise 0.
\end{itemize}
We have also included the ROC curve (receiver operating characteristic curve) which is a graph showing the performance of a classification model at all classification thresholds. Confusion matrices for the baseline and main models are included as well.

If we had more time, we were also going to look at another metric called Relative Improvement (RI) of AUC and Logloss, defined as,
\begin{align}
    RI-X = \dfrac{|X(model) - X(base)|}{X(base)}*100\%
\end{align}

Where $X$ is AUC or Logloss. This is a useful metric because a small improvement with respect to AUC is regarded significant for real-world CTR tasks. But, it cold be an insightful metric if we had used a larger/more complete version of this dataset. So, we planned to postpone it to future work. Even though there are a few qualitative metrics for evaluating recommender systems in general (we have mentioned some of them in the Future Work section), all of the CTR prediction literature we studied did suffice to AUC and Logloss, without really reporting a qualitative evaluation metric. That's because those qualitative metrics are more suited for a multi-label recommendation system e.g. Netflix's movie recommender, which has a little different nature from what we are trying to predict here. Nevertheless, we are reporting the recall, precision and F1-score on top of AUC and Logloss, in order to offer a complete analysis for our classification problem.

Moreover, we also tried to evaluate the effect of online\_learning\_to\_predict\_ratio on AUC value. 

\section{Results and Analysis}

There has been several interesting discoveries from our experiments. We initially set our batch size to 100, meaning that it learns after every 100 test input. Surprisingly, the resultant AUC actually went down to 0.7273. We explored further by increasing the learning batch size to 200. To our delight, the result actually went up to 0.7495 AUC. This observation challenges the naive view that the more frequently we update our model, the better it tracks the test input. Instead, if we update our model too closely to the test inputs, our training becomes greedy and the model become overfitted with the previous test input, resulting in worse model performance. 

\begin{figure}[htbp]
\centering 
\includegraphics[width=\columnwidth]{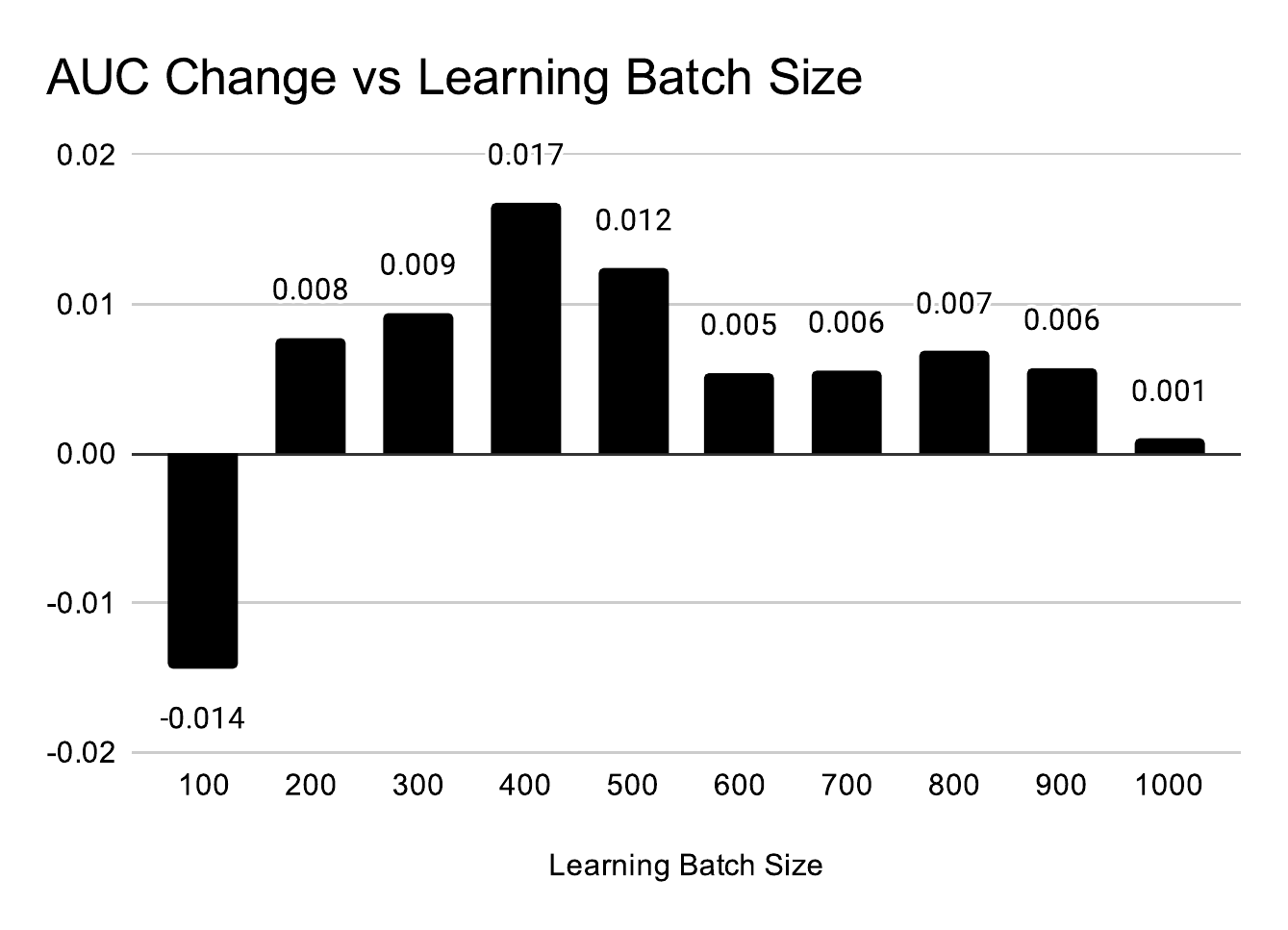} 
\caption{AUC vs learning batch size}
\label{fig:auc_vs_learning_batch}
\end{figure}

However, if the online learning batch size goes to high, it will also hinder our model's online performance. The larger the online learning batch size, the fewer learning iteration would the model go through during online operations. As result, the model wouldn't learn very frequently and may not be able to capture some sudden changes in user preferences. Moreover, since we have a limited size test input of 1000 data points, a large online learning batch size might leave no further ADs input after just a few learning batches. For example, when the batch size goes to 500 and above, the model will only be retrained once before exhausting all test input. This is clearly shown in Figure \ref{fig:auc_vs_learning_batch} and Table \ref{tbl:batch_size_auc}, which shows the changes in AUC with respect to the baseline result mentioned in earlier section. From the results, we can see that the AUC is actually maxed out at 400 batch size and we improved it by about RI-AUC= $1.7\%$. 

\begin{table}[htbp]
\centering
\begin{tabular}{|c|r|r|}
\hline
\textbf{Models}                                                                                            & \multicolumn{1}{c|}{\textbf{AUC}} & \multicolumn{1}{c|}{\textbf{Log loss}} \\ \hline
Baseline-DeepFM                                                                                            & \multicolumn{1}{c|}{0.6428}       & \multicolumn{1}{c|}{1.5371}            \\ \hline
\begin{tabular}[c]{@{}c@{}}Fi-GNN\\ (training batch size 1024,\\ no online learning)\end{tabular}          & \multicolumn{1}{c|}{0.4918}       & \multicolumn{1}{c|}{0.8142}            \\ \hline
\begin{tabular}[c]{@{}c@{}}Fi-GNN\\ (batch size 10, \\ no online learning)\end{tabular}                    & \multicolumn{1}{c|}{0.7533}       & \multicolumn{1}{c|}{0.4591}            \\ \hline
\begin{tabular}[c]{@{}c@{}}Baseline Fi-GNN\\ (batch size 100, \\ no online learning)\end{tabular}          & \multicolumn{1}{c|}{0.7418}       & \multicolumn{1}{c|}{0.4688}            \\ \hline
\begin{tabular}[c]{@{}c@{}}Fi-GNN + OL\\ (batch size 100, \\ online learning batch size 100)\end{tabular}  & 0.7273                            & 0.4939                                 \\ \hline
\begin{tabular}[c]{@{}c@{}}Fi-GNN + OL\\ (batch size 100, \\ online learning batch size 200)\end{tabular}  & 0.7495                            & 0.4722                                 \\ \hline
\begin{tabular}[c]{@{}c@{}}Fi-GNN + OL\\ (batch size 100, \\ online learning batch size 300)\end{tabular}  & 0.7512                            & 0.4622                                 \\ \hline
\begin{tabular}[c]{@{}c@{}}Fi-GNN + OL\\ (batch size 100, \\ online learning batch size 400)\end{tabular}  & 0.7585                            & 0.4666                                 \\ \hline
\begin{tabular}[c]{@{}c@{}}Fi-GNN +OL\\ (batch size 100, \\ online learning batch size 500)\end{tabular}   & 0.7542                            & 0.4589                                 \\ \hline
\begin{tabular}[c]{@{}c@{}}Fi-GNN + OL\\ (batch size 100, \\ online learning batch size 600)\end{tabular}  & 0.7472                            & 0.4696                                 \\ \hline
\begin{tabular}[c]{@{}c@{}}Fi-GNN + OL\\ (batch size 100, \\ online learning batch size 700)\end{tabular}  & 0.7473                            & 0.4916                                 \\ \hline
\begin{tabular}[c]{@{}c@{}}Fi-GNN + OL\\ (batch size 100, \\ online learning batch size 800)\end{tabular}  & 0.7486                            & 0.4711                                 \\ \hline
\begin{tabular}[c]{@{}c@{}}Fi-GNN + OL\\ (batch size 100, \\ online learning batch size 900)\end{tabular}  & 0.7475                            & 0.4730                                 \\ \hline
\begin{tabular}[c]{@{}c@{}}Fi-GNN + OL\\ (batch size 100, \\ online learning batch size 1000)\end{tabular} & 0.7427                            & 0.4639                                 \\ \hline
\end{tabular}
\caption{Results with Different Batch Sizes}
\label{tbl:batch_size_auc}
\end{table}

The improvement is still rather small because our offline model was trained on 8,000 data points, whereas we only have 1,000 test inputs. In reality, we could see more test inputs that will likely improve our online learning model's performance. 

\section{Error Analysis}

\begin{figure}[htbp]
\centering 
\includegraphics[width=\columnwidth]{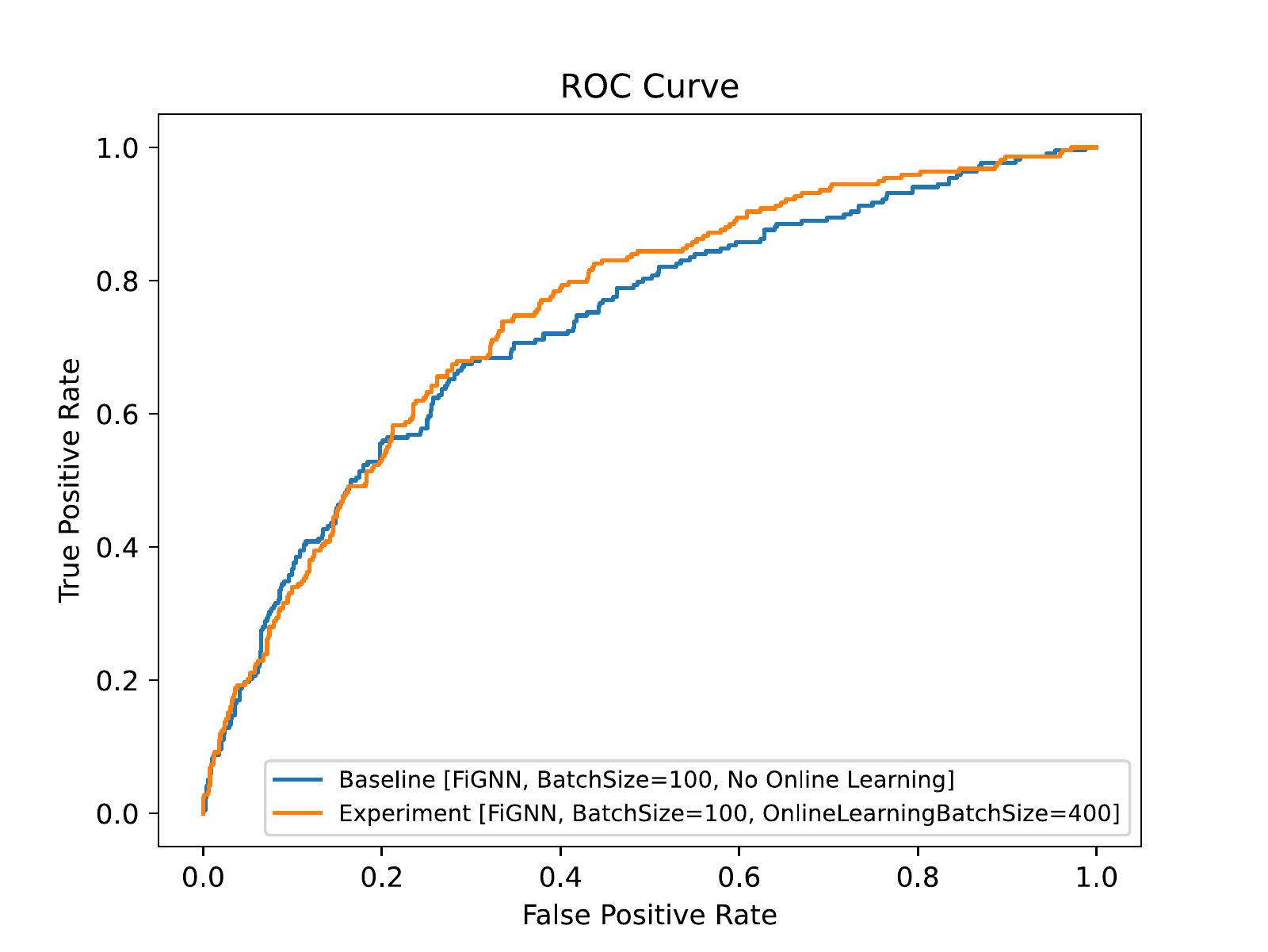} 
\caption{ROC Curve}
\label{fig:roc_curve}
\end{figure}

\begin{figure}[htbp]
\centering 
\includegraphics[width=\columnwidth]{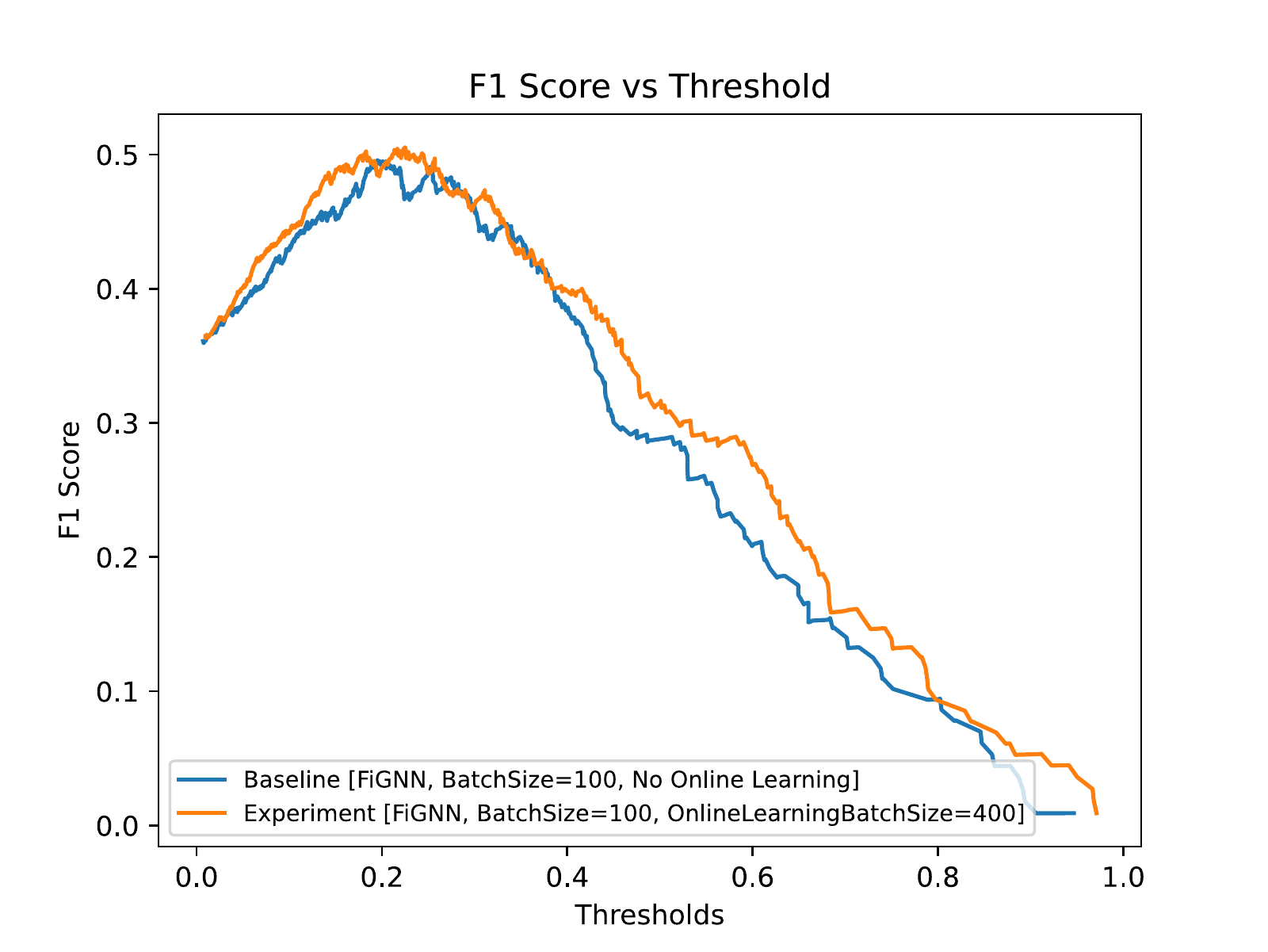} 
\caption{F1 Score vs Threshold}
\label{fig:f1_score}
\end{figure}

\begin{figure}[htbp]
\centering 
\includegraphics[width=\columnwidth]{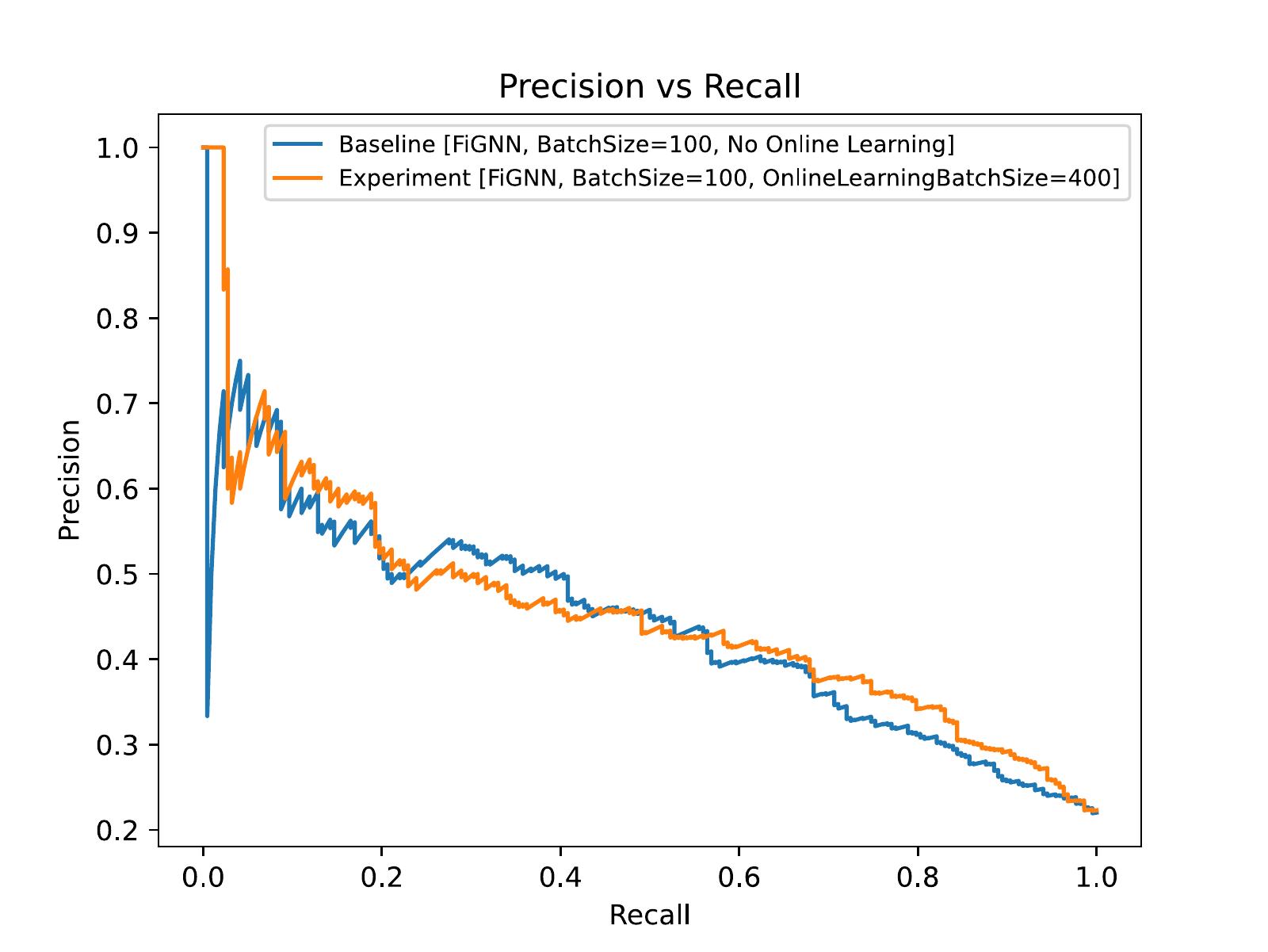} 
\caption{Precision vs Recall}
\label{fig:precision_recall}
\end{figure}

\begin{figure}[htbp]
\centering 
\includegraphics[width=\columnwidth]{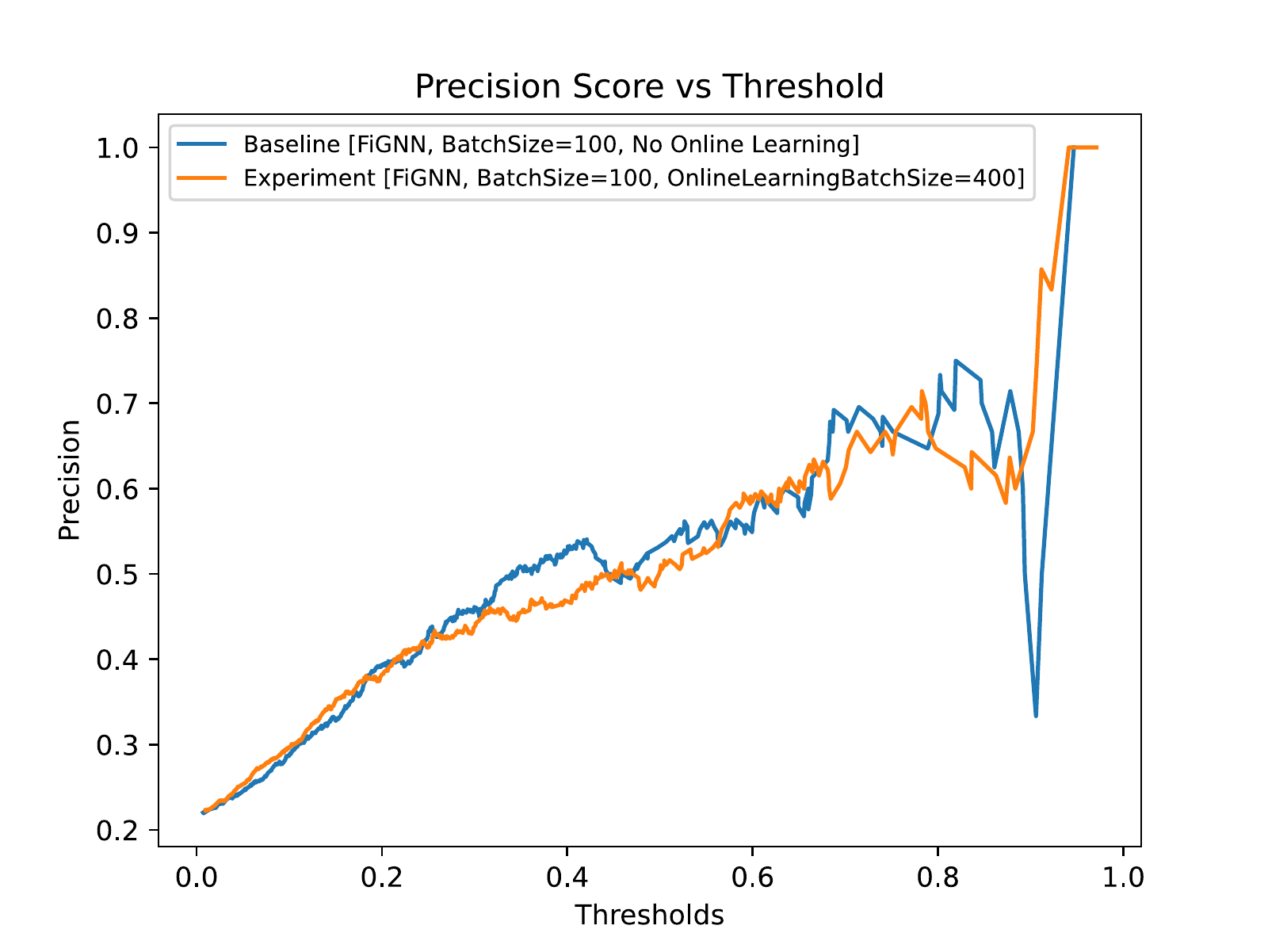} 
\caption{Precision Score vs Threshold}
\label{fig:precision_score}
\end{figure}

\begin{figure}[htbp]
\centering 
\includegraphics[width=\columnwidth]{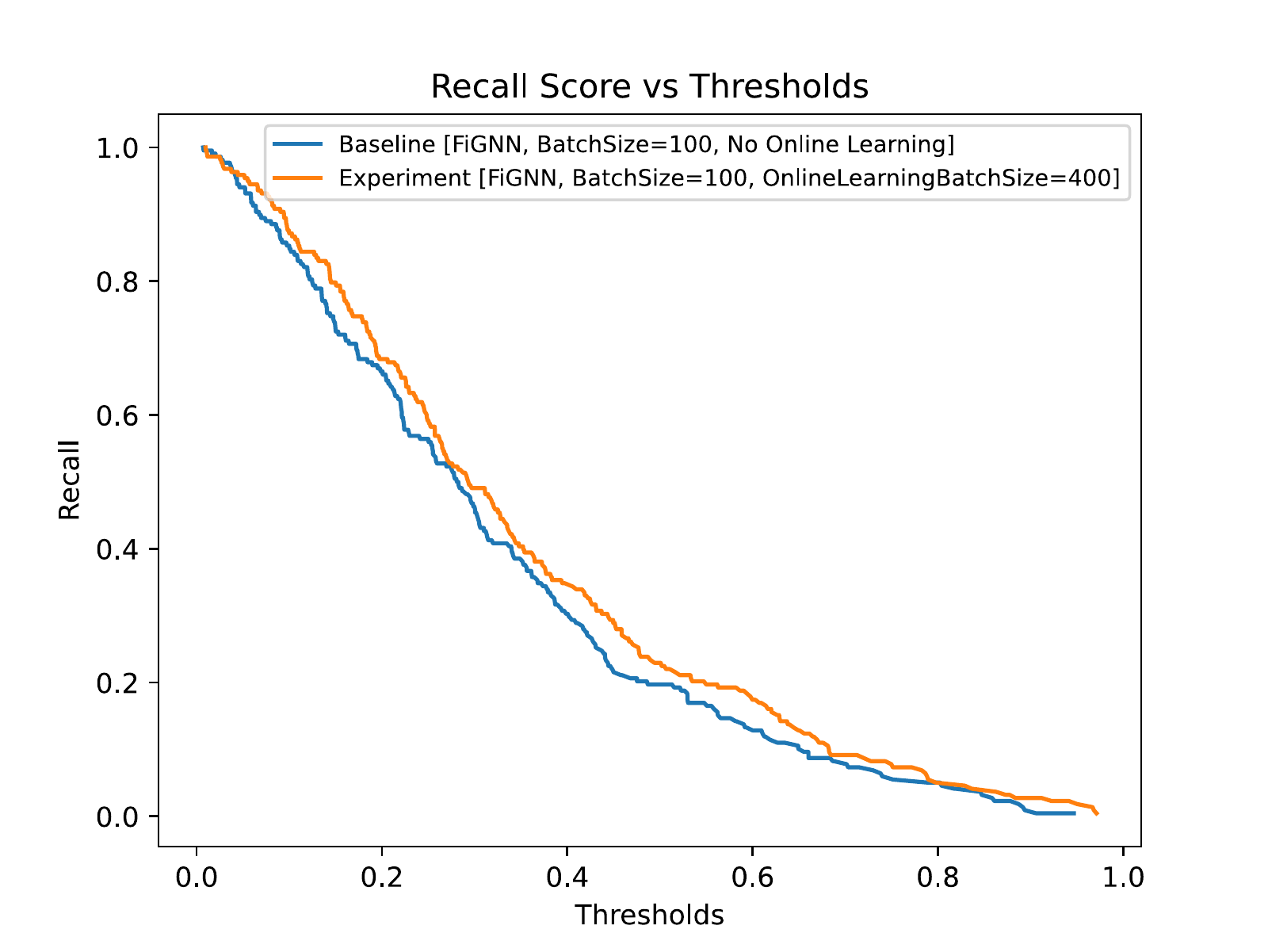} 
\caption{Recall Score vs Threshold}
\label{fig:recall_score}
\end{figure}

We have reported our analysis of this study both in terms of the AUC and Logloss values for every experiments we have performed. In Table \ref{tbl:batch_size_auc}, as well as the ROC curve (Figure \ref{fig:roc_curve}), F1-score curve (Figure \ref{fig:f1_score}) for different threshold values (the reported $\hat{y}$ from models is a probability, and thus converting it to a binary label to make it comparable to our true label requires a threshold value). We have also reported precision vs recall in Figure \ref{fig:precision_recall}, precision vs threshold in Figure \ref{fig:precision_score}, and recall vs threshold in Figure \ref{fig:recall_score}. All the reported results are the average of three runs for each case; because results may vary given the stochastic nature of the algorithm or evaluation procedure, or differences in numerical precision.

\begin{itemize}
    \item All of the plots correspond to two models, one baseline, and one baseline plus online batch learning. The baseline Fi-GNN model has batch-size 100. The Fi-GNN plus online learning model has batch size of 100 and learning batch size of 400. All results are reported for the optimum threshold value of 0.3 as represented by the F1-score vs threshold plot. Note that the two models being plotted have very close AUC, so they are not far from each other on the plots. If we had used the model with AUC of 0.4819 (second row in the table), we could clearly see higher improvements.
    \item According to the ROC curve, the experiment (Fi-GNN plus online learning) has a slightly higher area under curve than that of the baseline with no online learning, meaning that experiment has done a better job than the baseline in classifying the positive class in the dataset. However, as we will see in the confusion matrix, the experiment has done a perfect job in classifying the negative class, but has done poorly on classifying the positive class. That's because ROC curve is used when there are roughly equal numbers of observations for each class. Whereas, when there is a moderate to large class imbalance, Precision-Recall curve gives a better insight. The reason for this recommendation is that ROC curves present an optimistic picture of the model on skewed datasets. Therefore, the Precision-Recall Plot is more informative than the ROC Plot when evaluating binary classifiers on imbalanced datasets. We did investigate our dataset after running models, and it was indeed imbalanced, with the majority being negative class. This makes the classifier highly biased 
    
    \item The Precision vs. threshold  curve reveals that higher thresholds result in higher precisions and, high precision relates to a low false positive rate. The experiment model shows slightly higher false positive rates than the baseline, for most  thresholds.
    \item The Recall vs. threshold curve reveals that lower thresholds result in higher recall and, high recall relates to a low false negative rate. also, the online learning recall curve vs the baseline shows that online learning has slightly less false negative rate.
    \item The sweet spot between precision abd recall threshold values is where F1-score is maximized. We see from F1-score vs. threshold curve that optimal value of 0.3 maximizes F1-score. So, at a threshold of 0.3 for separating the probabilities of the positive class from negative class, F1-score has its maximum value. 
    \item According to the confusion matrices for both baseline and our online learning experiment, while the baseline model does a reasonable job in classifying the true positives and true negatives (Precision= 0.41, Recall=0.57, Accuracy=0.73, negative class accuracy=0.77), the online learning model does a perfect job on predicting negative class (96\% correct), and does a poor job on predicting the positive class (20\% correct). For this experiments we got Precision= 0.55, Recall=0.2, Accuracy=0.8, negative class accuracy=0.96. The high false negative rate again confirms how skewed our dataset is. The recall for this model is much lower than the baseline's. So, any improvements on this model (making the dataset balanced and then improving the model) could start by making changes and watching how recall is impacted.
\end{itemize}

\begin{figure}[htbp]
\centering 
\includegraphics[width=45mm]{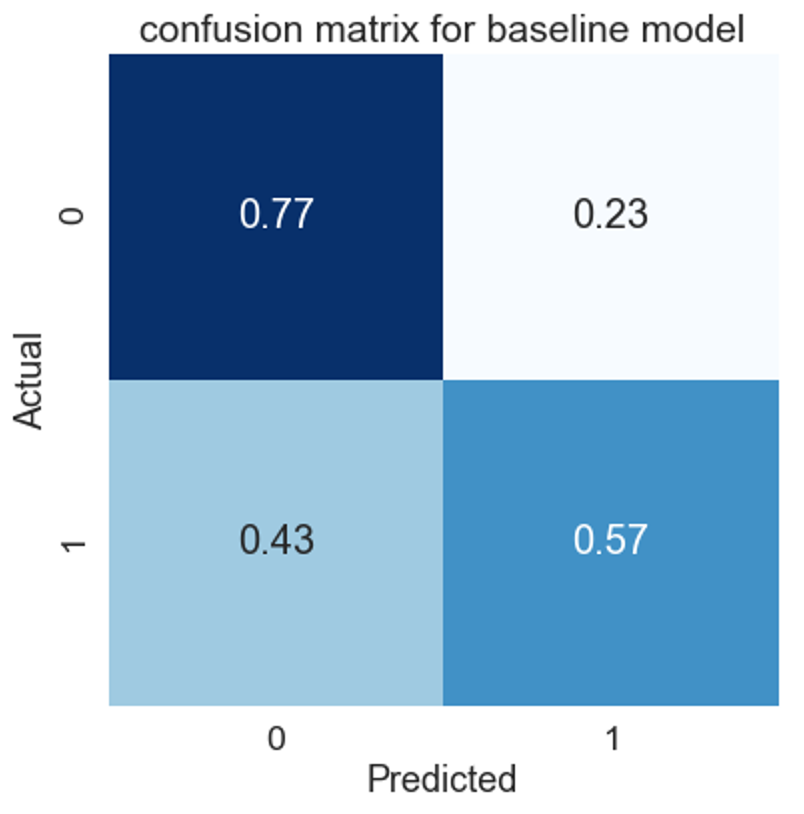} \\
\includegraphics[width=45mm]{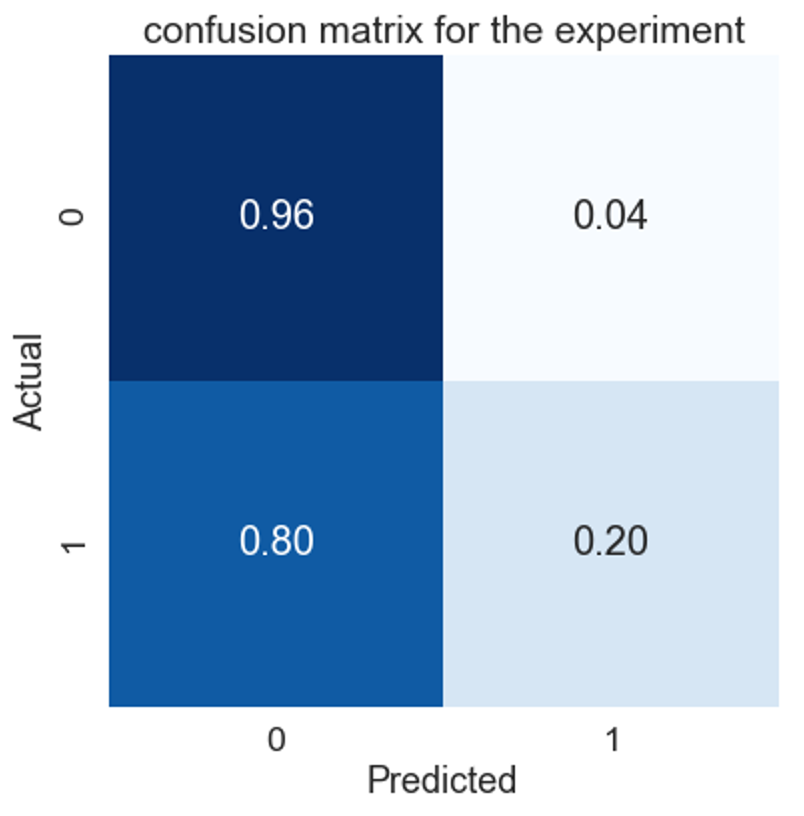} 
\caption{Confusion matrices for the baseline model and the best experiment, both with threshold 0.3 and training batch size of 100. The experiment matrix has a learning batch size of 400}
\label{fig:confusion matrix}
\end{figure}
\section{Future Work}

Due to the time constraint, we could not finish a lot of ideas that we had planned for this problem. If we had more time, we were going to finish our implementation of DRL approach proposed by \cite{liu2018deep}, and compare it against the online-learning algorithm we implemented on top of the Graph-interaction Network, as well as our other baseline Deep Factorization Machines which is still widely being used as a promising model in the practical/industrial applications. 

Also, we did not have enough time to perform the experiments on larger version of the Criteo dataset. We believe that using a larger version could extensively improve our reported accuracies. Also, we could not do much hyper-parameter tuning on our experiments. This is a major step in any ML model, but we just postponed it to a future attempt.

Inspecting data-imbalance was another point mentioned before. We did not check this here, but we believe it would considerably boost our accuracies if we make sure the dataset in not unreasonably imbalanced in terms of positive and negative classes.

One future direction could be building a recommendation engine using a powerful and efficient dynamic adaptation algorithm (like an Actor-Critic RL scheme), which is able to recommend the top-K items that users are most likely to click based on each user's history of clicking on the past recommended items. This is similar to how the Netflix's Movie recommendation algorithm does an offline testing, followed by an online A/B testing, resulting in the top-K movies to recommend \footnote{https://netflixtechblog.com/netflix-recommendations-beyond-the-5-stars-part-1-55838468f429}, \footnote{https://netflixtechblog.com/netflix-recommendations-beyond-the-5-stars-part-2-d9b96aa399f5}. Having such a top-K recommender, we could analyze a wider range of RS-specific quantitative and qualitative metrics like: top-N hit-rate, average reciprocal rank, cumulative hit-rate, rating hit-rate, coverage, diversity, novelty, churn, responsiveness,  and so on. Yet, besides all of these, the online A/B testing has been mentioned as the most reliable metric for making better recommendations.

To this end, by simplifying the problem into a CTR predictor, we did not have to deal with the K nearest neighbor pruning that can be technically challenging and was beyond our time limit. Nevertheless, by recasting the CTR problem into a multi-label binary classification approach (and applying an algorithm like Binary Relevance for the multi-label problem if there's no correlation between the different recommendations, or using a label powerset approach if we believe there is correlation between the recommended campaigns), and outputting the probability of the user clicking on a specific recommendation, followed by sorting all the outputted probabilities, and then picking the top-K (e.g. top-10), we can unambiguously extend our models into a top-K recommender system. 

Another idea we could try to improve AUC was performing an ablative study similar to the one in \cite{Li_2019}. Their idea was adding an edge-wise node interactions via attentional edge weights and edge-wise transformation, as well as extra residual connections to update state along with GRU in a GGNN. We could remove any of these additions (e.g. edge-wise interactions) and rerun the online learning to see whether we get any improvements. 

Another very interesting approach that we would like to investigate in the future, is implementing a Meta-learning approach \cite{cunha2018metalearning}, \cite{luo2020metaselector}. The motivation behind such model is that the recommender sytem's dataset are usually very heterogeneous and highly personalized, and no single model could give the best recommendation for every user. Therefore, a user-level adaptive model selection will train a collection of different recommendation models (including DFM, Fi-GNN, DRL, and so on) with data from all users, on top of which a model-selector is trained via meta-learning to select the best single model for each user with the user-specific historical data. Such an approach has been reported to achieve improvements over the  single-model baseline. 



\printbibliography
\vspace{12pt}

\end{document}